\documentclass[lettersize,journal]{IEEEtran}
\usepackage{stfloats}
\usepackage{url}
\usepackage{verbatim}
\usepackage{cite}
\usepackage{amsmath,amssymb,amsfonts}
\usepackage{algorithmic}
\usepackage{algorithm}
\usepackage{graphicx}
\usepackage{textcomp}
\usepackage{balance}
\usepackage[dvipsnames,table,xcdraw]{xcolor}
\usepackage[utf8]{inputenc}
\usepackage{booktabs}
\usepackage{multirow}
\usepackage{placeins}
\usepackage[caption=false, font={rm, scriptsize}]{subfig}
\usepackage{url}

\usepackage{array}
\newcolumntype{L}[1]{>{\raggedright\let\newline\\\arraybackslash\hspace{0pt}}m{#1}}
\newcolumntype{C}[1]{>{\centering\let\newline\\\arraybackslash\hspace{0pt}}m{#1}}
\newcolumntype{R}[1]{>{\raggedleft\let\newline\\\arraybackslash\hspace{0pt}}m{#1}}

\begin{document}

\title{
Economic Fairness of Operating Envelopes for Distributed Energy Resource Aggregations
}
\author{Noah Rhodes, Johanna L. Mathieu
\thanks{N. Rhodes is with Los Alamos National Laboratory, Los Alamos, NM, USA, {\tt nrhodes@lanl.gov}. J.L. Mathieu is with the Department of Electrical Engineering \& Computer Science, University of Michigan, Ann Arbor, MI, USA, {\tt jlmath@umich.edu}.}}

\maketitle

\begin{abstract}
Operating envelopes are used to protect distribution networks from uncoordinated power flows caused by distributed energy resources (DERs). Fairness is an important consideration; otherwise, DERs further from the substation are generally more constrained by the distribution network and are unable to inject or consume as much power as a customer near the substation, i.e., treated unfairly. There are several ways of incorporating fairness into the operating envelope problem and the literature has not standardized on one approach.  If DERs are participating as an aggregation in, e.g., electricity markets, we can decouple physical and economic fairness.  In this scenario, payments do not necessarily have to directly relate to physical constraints and power flow.  In this paper, we explore the value of decoupling physical and economic fairness, and different ways to allocate the ``fairness surplus" that arises when the physical problem is unfair.
We find that utilizing economic fairness can unlock significantly more value than physical fairness, but a lack of physical fairness can result in an energy service like space cooling being unavailable to a customer for a period of time.
These findings have implications for the design of operating envelopes and associated energy policy.
\end{abstract}


\section{Introduction}
More distributed energy resources (DERs) are being added to the grid each year, with distributed solar accounting for 42\% of all solar PV expansion worldwide in 2025 \cite{iea2025renewables}.
DERs include distributed generation like solar PV, as well as batteries and flexible electric loads such as electric vehicles. 
When a large number of DERs are operating on the same feeder it can lead to problems such as under/over-voltages and transformer overloading \cite{ross2018effects, hashemi2017methods, ul2015impact}.
There are a variety of proposed approaches to addressing this challenge. One is referred to as ``dynamic operating envelopes" \cite{blackhall2020calculation, Petrou2021, Liu2021, liu2022using}, or simply ``operating envelopes." Operating envelopes provide limits on power injection and consumption at each bus to maintain network voltage and current limits. These limits are a function of network loading, and are ``dynamic" or change over time to reflect the state of the system.
Operating envelopes are used in Australia to mitigate the impact of solar PV on distribution networks. This results in solar PV occasionally being curtailed to prevent transformer thermal limits and voltage limits from being exceeded \cite{arena2022}.

Fairness is an important consideration when using operating envelopes, e.g., see~\cite{soares2024review,yi2022fair,gebbran2021fair,liu2020fairness,petrou2020operating}. If fairness is not considered, DERs physically closer to the substations typically will typically be given more flexibility, since remote buses generally hit voltage limits first. 
There are many different ways to define fairness, and the application of fairness has been widely studied particularly in the communication networks and machine learning literatures~\cite{mo2002fair, paganini2005congestion, srivastava2019mathematical, mehrabi2021survey}. Many of these fairness metrics can be used in optimization as a constraint or by adding the metric into the objective function \cite{xinying2023guide}. As shown in approaches that impose fairness in operating envelope calculations~\cite{yi2022fair,liu2020fairness,petrou2020operating,moring_fair_2024}, imposing fairness generally reduces flexibility, which presents a tradeoff.

Usually operating envelopes are proposed in settings where DERs are otherwise uncoordinated, e.g., to manage solar PV injections across a feeder. However, there are now a variety of applications where DERs are actively coordinated/controlled by an entity such as a DER aggregator. For example, DER aggregators can participate in US electricity markets per FERC order 2222 \cite{FERC2020Order2222}. 
Some work has proposed using operating envelopes to curtail the actions of DER aggregators \cite{liu2022using, russell2024network, moring2024nodal}. In that case, a utility company (distribution system operator) would compute operating envelopes and send them to the DER aggregator, which would design its coordination/control signals to ensure DER injections/consumption remain within their operating envelopes.
In this case, since physical injections are coordinated, and since incentives and payments are also coordinated by the DER aggregator, we have the opportunity to decouple physical fairness (i.e., fairness of the injection limits) and economic fairness (fairness of the incentives/payments), which are otherwise linked. An aggregator can physically operate the DERs unfairly to maximize revenue, and then allocate (some of) its profit back to the DER owners ``fairly", possibly in a way not directly reflecting their individual real-time participation. This paper investigates this approach. 

The contributions of this work are as follows. 
1) We develop a framework for decoupling physical and economic fairness in the operating envelope problem relevant for DER aggregators, and we define the concept of the ``fairness surplus".
2) We provide a comparison of different methods of imposing fairness in the operating envelope problem, and methods of distributing the ``fairness surplus".
3) We use a case study to demonstrate the benefits and drawbacks of physical versus economic fairness within the operating envelope problem, and different variants of the economic fairness approach.
4) Finally, we discuss the implications of our results for the design of operating envelopes and associated energy policy. 
Our results will help power systems better leverage DERs for power system market participation and services in a way that ensure safety and power quality in the distribution network.

This paper is organized as follows. Section~\ref{sec:setting} describes the problem setting. Section~\ref{sec:physical_models} introduces the operating envelope optimization models for physical operation and Section~\ref{sec:economic_fairness} introduces methods for distributing the economic value.
Section~\ref{sec:case_study} details the case study and Section~\ref{sec:results} presents the results.  Finally, we discuss the implications of the results in Section~\ref{sec:discussion} and Section~\ref{sec:conclusion} concludes the work.

\section{Problem Setting}
\label{sec:setting}
Operating envelopes are usually applied in a setting without an aggregator, e.g., curtailment of residential PV by the utility. 
In a setting with aggregators, DERs work together for some goal.
The aggregator receives a financial incentive to coordinate the DERs and then compensates the owner/operator of DERs within the aggregation. That provides an opportunity -- aggregators could enable sharing the (financial) impact of
operating envelopes. 

\begin{figure}[t]
    \centering\includegraphics[width=0.65\linewidth]{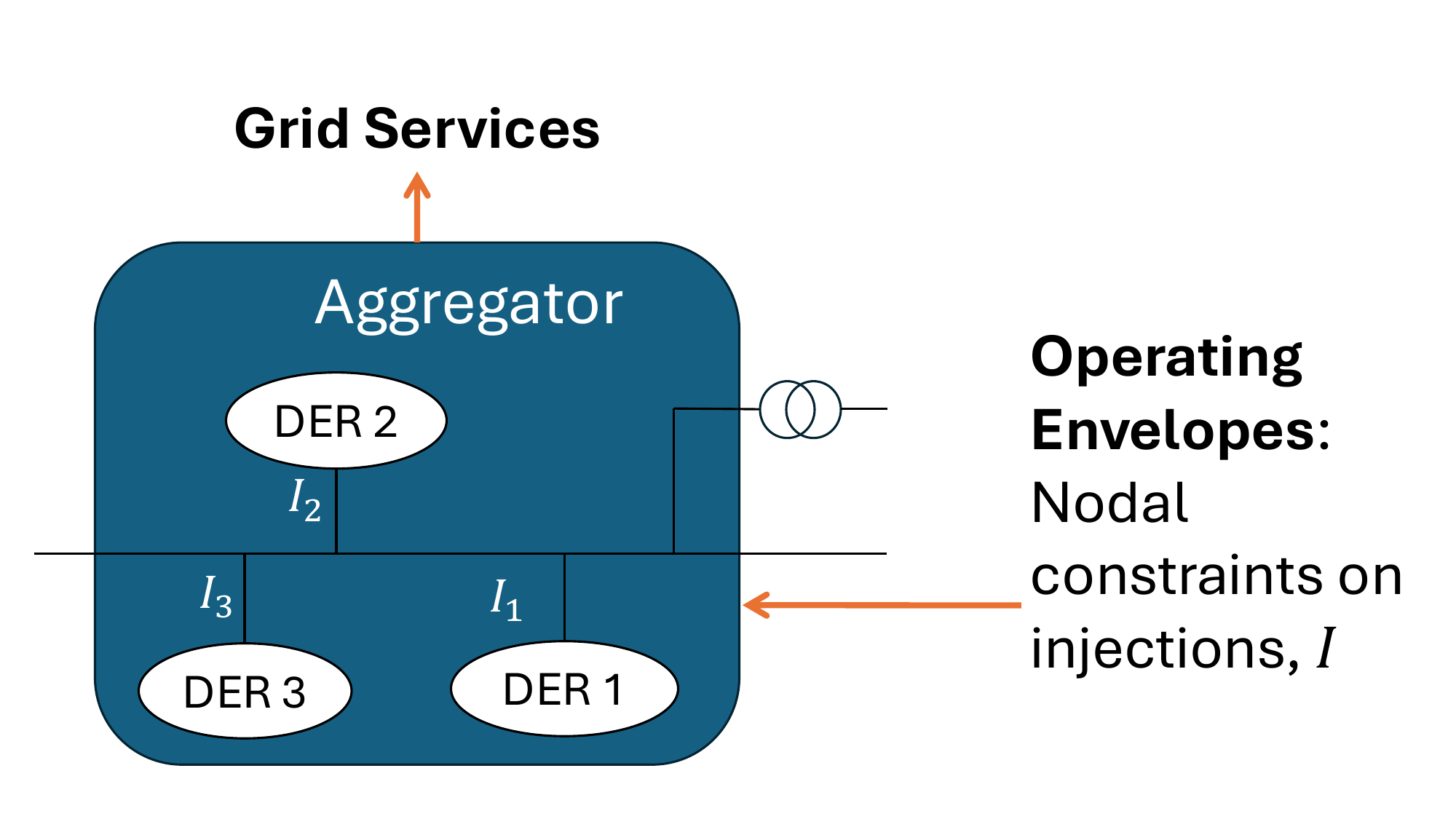}
    \caption{Operating envelopes imposed on a DER aggregator.}
    \label{fig:aggregator}
\end{figure}
\begin{figure}[t]
    \centering
    \includegraphics[width=0.6\linewidth]{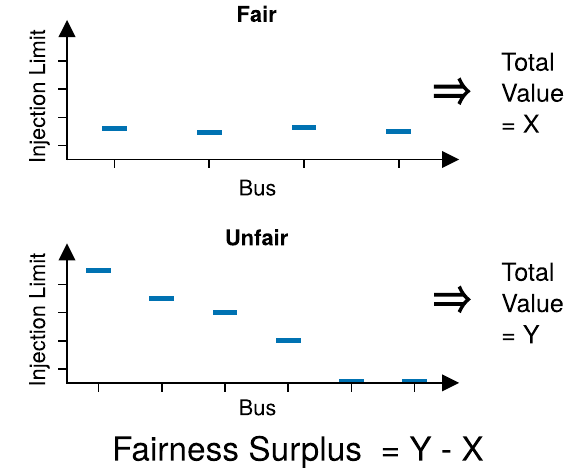}
    \caption{The fairness surplus is the difference in value (flexibility of the aggregation constrained by operating envelopes) of an unfair and fair allocation of operating envelopes.}
    \label{fig:fairness_surplus}
    \vspace{-1.5em}
\end{figure}

Fig.~\ref{fig:aggregator} shows a setting where an aggregator coordinates DERs for wholesale electricity market participation. 
The aggregator wants to maximize their profit (revenue minus cost). 
Operating envelopes can reduce revenue if they require aggregators to reduce the flexibility they offer to the market. Imposing fairness further reduces flexibility. Instead, the aggregator could maximize flexibility (i.e., use operating envelopes without considering fairness) and then compensate the DERs within its aggregation fairly. This decouples physical fairness and economic fairness. The approach is physically unfair but economically fair. 

We refer to the surplus achieved when maximizing flexibility versus imposing fairness as the ``fairness surplus", with an example shown in Fig.~\ref{fig:fairness_surplus}. There are many options for how to allocate the surplus, and
it is not clear how to best design the compensation and what that implies for physical and economic performance. This paper investigates the impact of decoupling physical and economic performance, and examines several different implementations of physical fairness and economic fairness.
Our results have implications for operating envelope design and associated energy policy design.

\section{Operating Envelope Optimization Models}
\label{sec:physical_models}
We first introduce the operating envelope optimization problem with AC power flow and an objective that maximizes the total size of the envelopes.  We then introduce the optimization problems for operating envelopes with Minimax fairness, $\alpha$-fairness, and a Lexicographic objective function \cite{rentmeesters1996theory}.  

\textbf{Utilitarian Operating Envelope:}
We consider a single phase equivalent balanced distribution network with buses $i \in \mathcal{B}^\mathrm{all}$ and lines $(ij) \in \mathcal{L}$. The notation $\mathcal{L}_i$ indicates the set of lines that connect to bus $i$.
The bus at the substation is considered a reference bus $i \in \mathcal{B}^\mathrm{ref}$, while the set of non-substation buses is  $\mathcal{B} := \mathcal{B}^\mathrm{all} \setminus \mathcal{B}^\mathrm{ref}$.  In the following, bold symbols are parameters while non-bolded symbols are variables.

The total active and reactive power injections from all DERs connected to bus $i$ are denoted $P^{C}_i$ and $Q^{C}_i$. They are physically limited, i.e.,
\begin{subequations}
    \begin{align}
        \underline{\boldsymbol{P}}^{C}_i \le  P^{C}_i \le \overline{\boldsymbol{P}}^{C}_i & \quad \forall i \in \mathcal{B}^\mathrm{all}, \\
        \underline{\boldsymbol{Q}}^{C}_i \le Q^{C}_i \le \overline{\boldsymbol{Q}}^{C}_i  & \quad\forall i \in \mathcal{B}^\mathrm{all} .
    \end{align} \label{eq:bus_power_limits}%
\end{subequations}
where $\overline{\boldsymbol{P}}^{C}_i$, $\overline{\boldsymbol{Q}}^{C}_i$ and $\underline{\boldsymbol{P}}^{C}_i$, $\underline{\boldsymbol{Q}}^{C}_i$ are the sum of the upper and lower physical limits of all DERs connected to bus $i$.

The active and reactive power balance equations at each non-substation bus are
\begin{subequations}
    \begin{align}
         P_i^{C}  - \!\!\!\!\!\! \sum_{(i j)\in\mathcal{L}_i} \!\!\!\!\!  P_{i j}^L - \! \boldsymbol{P}^D_i  = 0 & \quad  \forall i \in \mathcal{B} ,\label{eq:ops_active_power_balance} %
        \\
        Q_i^{C} - \!\!\!\!\!\! \sum_{(i j)\in\mathcal{L}_i} \!\!\!\!\!  Q_{i j}^L - \! \boldsymbol{Q}^D_i = 0 & \quad \forall i \in \mathcal{B}, \label{eq:ops_reactive_power_balance}  %
    \end{align} \label{eq:power_balance}%
\end{subequations}
where $P^L_{ij}, Q^L_{ij}$ are the active and reactive power flows on line $(ij)$ and $\boldsymbol{P}^D_i, \boldsymbol{Q}^D_i$ are the total active and reactive power consumed by all loads connected to bus $i$.   

At the substation bus there is no need for an operating envelope since its voltage is regulated. Let $P^S_i, Q^S_i$ be the amount of power the substation draws from the transmission network. Then the active and reactive power balance equations  at the substation are
\begin{subequations}
    \begin{align}
         P_i^S+P_i^{C} - \!\!\!\! \sum_{(i j)\in\mathcal{L}_i} \!\!\!\!  P_{i j}^L - \! \boldsymbol{P}^D_i  = 0 & \quad  \forall i \in \mathcal{B}^\mathrm{ref} , \label{eq:ops_active_power_balance_substation} %
        \\
        Q_i^{S}+Q_i^{C} \! - \!\!\!\! \sum_{(i j)\in\mathcal{L}_i} \!\!\!\!  Q_{i j}^L - \! \boldsymbol{Q}^D_i = 0 & \quad \forall i \in \mathcal{B}^\mathrm{ref}  \label{eq:ops_reactive_power_balance_substation} . %
    \end{align} \label{eq:power_balance_substation}%
\end{subequations}
The remaining constraints for substation voltage, bus voltage limits, line thermal limits, and AC power flow are standard and can be found in the appendix.  

Finally, the objective function is to maximize the utility of the operating envelopes of the DERs, i.e., the sum of the envelopes across all non-substation buses without any consideration of fairness, i.e.,
\begin{equation}
     \sum_{i\in \mathcal{B}} {P_i^{C} } \label{eq:obj_unfair} .
\end{equation}
The full optimization problem for the maximal injection Utilitarian operating envelope problem is
    \begin{align}
        & \max_{P^C, Q^C} \eqref{eq:obj_unfair} \quad \tag{OE$^{\text{U}}_{\text{max}}$} \\
        & \text{s.t.} \quad  \eqref{eq:bus_power_limits}-\eqref{eq:power_balance_substation}, \eqref{eq:voltage_reference}-\eqref{eq:ac_powerflow} \nonumber ,
    \end{align} \label{oe_max_sw}%
where \eqref{eq:voltage_reference}-\eqref{eq:ac_powerflow} are the standard AC power flow equations and engineering limits, which can be found in the appendix, and $P^C$ and $Q^C$ are vectors that contain the variables $P^C_i~\forall i \in \mathcal{B}$ and $Q^C_i~\forall i \in \mathcal{B}$ respectively. 
The same problem is solved as a minimization problem to find the maximal power draw of loads,
    \begin{align}
        & \min_{P^C, Q^C} \eqref{eq:obj_unfair} \quad \tag{OE$^{\text{U}}_{\text{min}}$} \\
        & \text{s.t.} \quad  \eqref{eq:bus_power_limits}-\eqref{eq:power_balance_substation}, \eqref{eq:voltage_reference}-\eqref{eq:ac_powerflow} \nonumber .
    \end{align} \label{oe_max_sw}%
We note that both OE$^{\text{U}}_{\text{max}}$ and OE$^{\text{U}}_{\text{min}}$ are nonconvex optimization problems and so there is no guarantee that injections {\em within} the region defined by these envelopes will be feasible (or that the solutions we find are optimal). Using this formulation requires a monotonicity assumption, i.e., that the relationship between bus power injections and network voltages is monotonic~\cite{moring2024nodal, moring_exploring_2026}. This assumption does not always hold~\cite{Liu2024Robust}, but often does empirically in radial distribution networks. Other operating envelope approaches have used power flow linearizations~\cite{Liu2024Robust,liu2022using,Petrou2021,Russell2023Robust,Attarha2024Shaped}, convex relaxations~\cite{yi2022fair}, or convex restrictions~\cite{Nazir2022,Liu2024Robust} to define convex operating envelope problems. However, each approach comes with advantages and disadvantages~\cite{moring_exploring_2026}. In this paper, we focus on fairness, not the question of what is the ``best" approach to incorporate power flow. We use the nonconvex AC power flow equations and we assume that injections within the region defined by the envelopes will be feasible.

\textbf{Minimax Fair Operating Envelope:}
The Minimax fairness objective maximizes the smallest operating envelope \cite{yi2022fair}.  This is the `most' fair solution in that it ensures each bus receives at least that size operating envelope.  

We introduce a new variable $\gamma$ that is less than or equal to every operating envelope,
\begin{equation}
    \gamma \leq P_i^{C} \quad \forall i \in \mathcal{B} \label{eq:size_fairness_max} .
\end{equation}
Then the objective is to maximize this lower bound,
    \begin{align}
        & \max_{P^C, Q^C, \gamma} \gamma \quad \tag{OE$^{\text{MM}}_{\text{max}}$} \\
        & \text{s.t.} \quad  \eqref{eq:bus_power_limits}-\eqref{eq:power_balance_substation}, \eqref{eq:size_fairness_max}, \eqref{eq:voltage_reference}-\eqref{eq:ac_powerflow} \nonumber .
    \end{align} \label{oe_max_mm}%
For the minimization problem, the inequality in \eqref{eq:size_fairness_max} must be flipped so that $\gamma$ upper bounds the bus power injections,
\begin{equation}
    \gamma \geq P_i^{C} \quad \forall i \in \mathcal{B} \label{eq:size_fairness_min} ,
\end{equation}
resulting in 
the Minimax fair minimization problem,
    \begin{align}
        & \min_{P^C, Q^C, \gamma} \gamma \quad \tag{OE$^{\text{MM}}_{\text{min}}$} \\
        & \text{s.t.} \quad  \eqref{eq:bus_power_limits}-\eqref{eq:power_balance_substation}, \eqref{eq:size_fairness_min}, \eqref{eq:voltage_reference}-\eqref{eq:ac_powerflow} \nonumber .
    \end{align} \label{oe_min_mm}%
\textbf{$\alpha$ Fair Operating Envelope:}
$\alpha$ fairness is a generalization of Minimax fairness ($\boldsymbol{\alpha}=\infty$), proportional fairness ($\boldsymbol{\alpha}=1$), and a Utilitarian solution ($\boldsymbol{\alpha}=0$).  A value of $\boldsymbol{\alpha}$ between 0 and $\infty$ finds a tradeoff between Minimax and Utilitarian solutions.  The $\alpha$ fairness definition from \cite{mo2002fair} is
\begin{equation}
    W_{\boldsymbol{\alpha}}(u) = \begin{cases}
    \frac{1}{1-\boldsymbol{\alpha}} \sum_i u_i^{1-\boldsymbol{\alpha}} & \boldsymbol{\alpha} \ge 0, \alpha \neq 1, \\
    \sum_i \log u_i & \boldsymbol{\alpha}=1 ,
\end{cases}    
\end{equation}
where $u$ is an arbitrary vector with elements $u_i$.

The $\alpha$ fairness function is applied to the vector of operating envelope values, resulting in the $\alpha$ fair 
problem,
    \begin{align}
        & \max_{P^C, Q^C} W_{\boldsymbol{\alpha}}(P^{C}) \quad \tag{OE$^{\text{A}}_{\text{max}}$} \\
        & \text{s.t.} \quad  \eqref{eq:bus_power_limits}-\eqref{eq:power_balance_substation}, \eqref{eq:voltage_reference}-\eqref{eq:ac_powerflow} \nonumber .
    \end{align} \label{oe_max_a}%
To model the minimum operating envelope problem with $\alpha$ fairness for net loads, we multiply $P^{C}$ by $-1$ to change the sign for negative power injection (i.e., net load) to a positive number. This is necessary because $W_{\boldsymbol{\alpha}}(u)$ is only 
defined for $u_i \ge 0$. Specifically, the problem becomes
    \begin{align}
        & \max_{P^C, Q^C} W_{\boldsymbol{\alpha}}(-P^{C}) \quad \tag{OE$^{\text{A}}_{\text{min}}$} \\
        & \text{s.t.}  \quad \eqref{eq:bus_power_limits}-\eqref{eq:power_balance_substation}, \eqref{eq:voltage_reference}-\eqref{eq:ac_powerflow} \nonumber .
    \end{align} \label{oe_min_a}%
Because the elements of $P^{C}$ must all have the same sign, it is not possible to consider networks that have some buses that are net injectors and some buses that are net consumers.  While it is technically possible to combine them if their respective bounds include $0$, we find this results in numerical issues when used in the optimization problem.

\textbf{Lexicographic Fair Operating Envelope:}
The Minimax fairness solution can leave additional network capacity unused if the excess network capacity does not increase the smallest operating envelope.  This can occur, for example, if one lateral in the distribution network is more constrained than another.

To utilize the full capacity of the network, but also achieve the Minimax fairness for the lowest capacity bus, we use a Lexicographic multi-objective method \cite{rentmeesters1996theory}.  We first solve the Minimax problem OE$^{\text{MM}}_{\text{max}}$ to find $\gamma^*$.  Next, we solve the Utilitarian problem to find the maximal usage of the network, subject to the constraint $\gamma = \gamma^*$ which ensures Minimax fairness of the solution.
Specifically, the second stage of the lexicographic optimization, which maximizes the Utilitarian objective while preserving the optimal Minimax fairness, is formulated as:
\begin{align}
    & \max_{P^C, Q^C, \gamma} \eqref{eq:obj_unfair} \quad \tag{OE$^{\text{LX}}_{\text{max}}$} \label{oe_max_lx}\\
    & \text{s.t.} \quad \gamma = \gamma^*, \label{eq:lex_gamma_lock_max} \\
    & \quad \quad \eqref{eq:bus_power_limits}-\eqref{eq:power_balance_substation}, \eqref{eq:size_fairness_max}, \eqref{eq:voltage_reference}-\eqref{eq:ac_powerflow}. \nonumber
\end{align} %
This can be similarly solved for the minimization 
problem,
\begin{align}
    & \min_{P^C, Q^C, \gamma} \eqref{eq:obj_unfair} \quad \tag{OE$^{\text{LX}}_{\text{min}}$} \label{oe_min_lx} \\
    & \text{s.t.} 
     \quad \eqref{eq:bus_power_limits}-\eqref{eq:power_balance_substation}, \eqref{eq:size_fairness_min},\eqref{eq:lex_gamma_lock_max},  \eqref{eq:voltage_reference}-\eqref{eq:ac_powerflow}. \nonumber
\end{align}

\section{Economic Distribution of Value}
\label{sec:economic_fairness}
We explore four methods of distributing the value of flexibility to members in an aggregation. The first three methods do not use the fairness surplus. The final method uses the fairness surplus when distributing the value to the members.

We treat the value of the operating envelope as proportional to size of the operating envelope.  For example, if we solve for the maximum and minimum Utilitarian operating envelopes with OE$^{\text{U}}_{\text{max}}$ and OE$^{\text{U}}_{\text{min}}$ and denote the optimal operating envelopes $P^{C,\mathrm{max*}}_i$ and $P^{C,\mathrm{min*}}_i$, then the size of the operating envelope for each bus is $E_i = |P^{C,\mathrm{max*}}_i -P^{C,\mathrm{min*}}_i|$.  At each bus $i$, the value of the flexibility allowed by the operating envelope is $v_i = \phi E_i$ and the total value is $V=\sum_{i\in\mathcal{B}} \phi E_i$.  In the following  definitions we assume $\phi=1$ for simplicity.

\textbf{Utilitarian:}
The Utilitarian distribution of value states that the DERs at each bus are  compensated according to the size of their bus's operating envelope,
$$v_i = E_i \quad \forall i \in \mathcal{B}.$$
Some buses may be allocated their full DER capacity while others may be allocated no capacity, i.e., an operating envelope of 0, and therefore not be allocated any value.  If a more fair objective is used to calculate the operating envelopes, the distribution of value will be more fair.

\textbf{Equality:}
An equal distribution of value ensures that the DERs at each bus are paid the same value for participating in an aggregation regardless of their capacity or the size of their bus's operating envelope,
$$ v_i = \frac{\sum_{j\in \mathcal{B}} E_j}{|\mathcal{B}|}  \quad \forall i \in \mathcal{B}.$$
In this case, resources are rewarded for participation and not penalized due to restricted network capacity (which they have no control over); however, small and large resources are paid the same amount.

\textbf{Capacity:}
The value of the operating envelopes can also be distributed according to the capacity of the DERs in the aggregation.  The proportion of the a bus's DER capacity relative to the total DER capacity is multiplied by the total size of the operating envelopes in the aggregation,
$$ v_i = \frac{| \overline{\boldsymbol{P}}^{C}_i - \underline{\boldsymbol{P}}^{C}_i |}{\sum_{j\in\mathcal{B}} | \overline{\boldsymbol{P}}^{C}_j - \underline{\boldsymbol{P}}^{C}_j | } \sum_{j\in \mathcal{B}} E_j  \quad \forall i \in \mathcal{B} ,$$
where, as in the Equality distribution, the value is distributed regardless of their operating envelope. In this case, resources at each bus are rewarded for participation as a function of their bus's capacity, but not penalized due to restricted network capacity.

\textbf{Baseline Fairness:}
Finally, we consider how to distribute the value of an operating envelope according to a baseline level of fairness and then distribute the remaining surplus in a utilitarian way. 
For example, operating a DER aggregation using a Utilitarian operating envelope maximizes the available flexibility; however, a Minimax fair operating envelope would ensure all buses with participating DERs receive the same operating envelope.  We consider the difference in value of these two operating envelopes as the fairness surplus, 
$$ S = \sum_{i\in\mathcal{B}} \left(E_i^\mathrm{Oper} - E_i^\mathrm{Base}\right) ,$$
as depicted in Fig.~\ref{fig:fairness_surplus}.
Here the operational operating envelope $E^\mathrm{Oper}$ (i.e., the one that is deployed) could be a Utilitarian envelope and the baseline operating envelope $E^\mathrm{Base}$ could be a Minimax fair envelope.
The Baseline Fairness distribution ensures that the DERs at each bus are compensated as if the baseline operating envelope $E^\mathrm{Base}$ was used, and the fairness surplus is distributed to the DERs at each bus proportionally to the size of the operating envelope in $E^\mathrm{Oper}$,
\begin{equation}
    v_i = E^\mathrm{Base}_i + \frac{E^\mathrm{Oper}_i}{\sum_{j\in\mathcal{B}}E^\mathrm{Oper}_j} S \quad \forall i \in \mathcal{B}.
\end{equation}
This case enables the aggregator to maximize flexibility but compensate participants more fairly.

\section{Case Study}
\label{sec:case_study}

We use two DER scenarios on a 56 bus single phase equivalent of the IEEE 123 distribution network \cite{bolognani2015existence}.  This network has been modified to introduce congestion that necessitates operating envelopes to restrict the operations of DERs.  The bus voltage limits are restricted to $[0.95, 1.05]$ and the substation voltage is set to $1.02$ p.u. The six buses closest to the substation have approximately 10x higher load (and DER capacity) representing commercial customers that are closer to the substation. This also allows us to demonstrate the effects of imposing fairness on DERs of various capacities.

\begin{figure}[t]
    \centering
    \includegraphics[width=\linewidth]{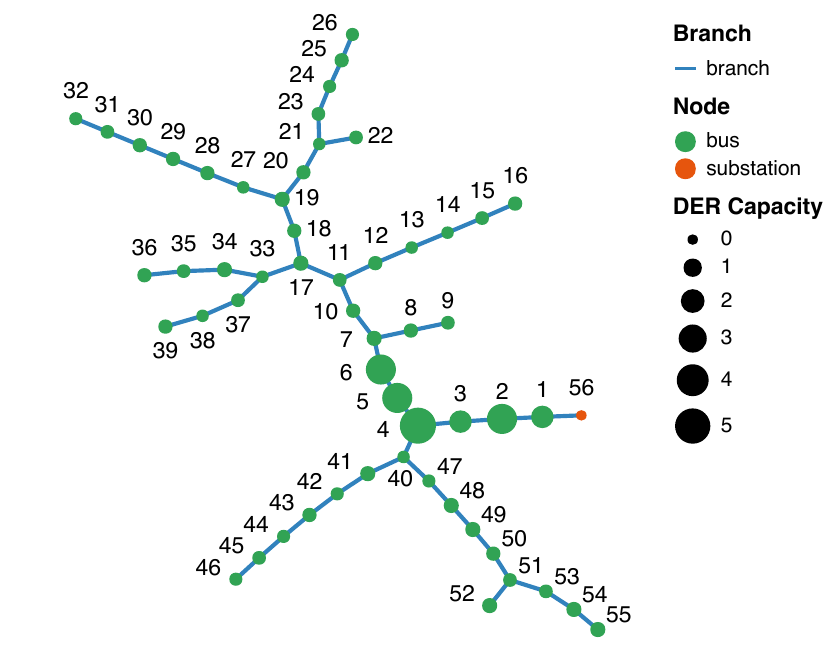}
    \vspace{-1em}
    \caption{Test Network under DG Scenario.  Buses are labeled with an id number and all buses have DG except the substation bus (56).}
    \label{fig:pv_scenario}
    \vspace{-1.0em}
\end{figure}

One scenario includes Distributed Generation (DG) at all buses in the network (except the substation), visualized  in Fig.~\ref{fig:pv_scenario} using PowerPlots.jl~\cite{rhodes2025powerplots}, where the total capacity of injected power would cause the network to exceed its limits, requiring operating envelopes that result in curtailment of the generation. The second scenario includes Flexible Load (FL) at most buses, which increases the loading of the network and requires limits on the power consumption to ensure bus voltage limits are not exceeded.  The relative scale of FL power at each bus is comparable to the DG scenario. The data and code are available on GitHub \cite{FairDERs}.

\section{Results}
\label{sec:results}
We first solve for the operating envelopes with each of the fairness objectives and observe the impact on each of the buses.  We then compare the value of the operating envelopes and how the value is distributed according to the different methods of fairness.

\subsection{Operating Envelopes with Fairness}

\begin{figure*}[t]
    \centering
    
    \subfloat[Maximum operating envelope for distributed generation at each bus.\label{fig:pv_oe_max_barplot}]{
        \includegraphics[width=0.95\linewidth]{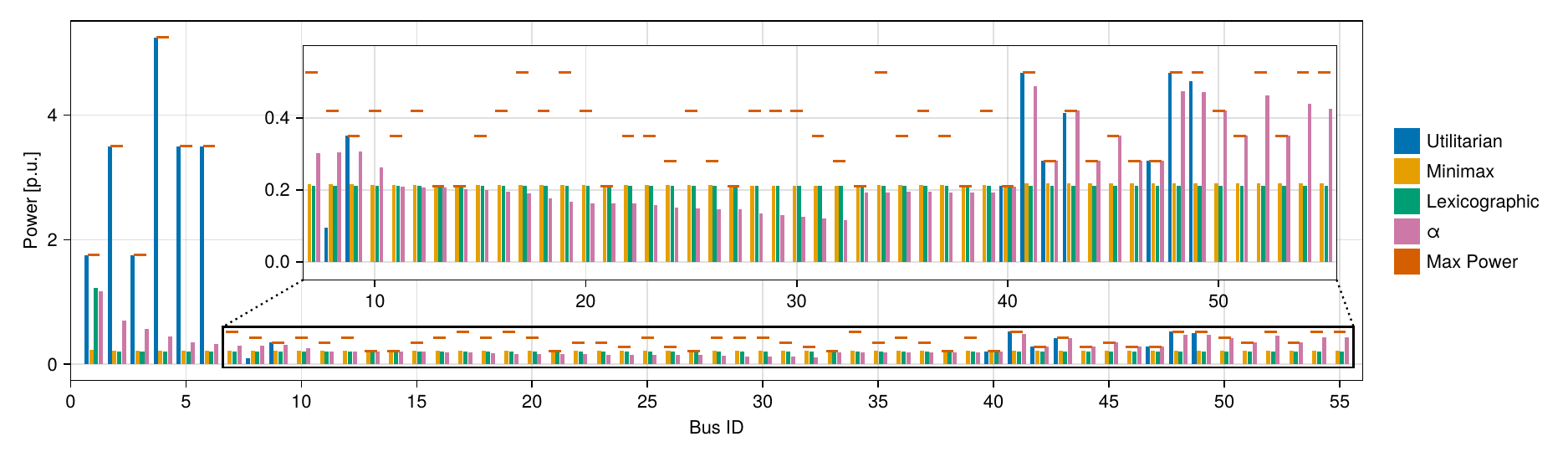}
    } \\
    \vspace{-0.5em} %
    
    \subfloat[Minimum operating envelope for flexible load at each bus.\label{fig:tcl_oe_min_barplot}]{
        \includegraphics[width=0.95\linewidth]{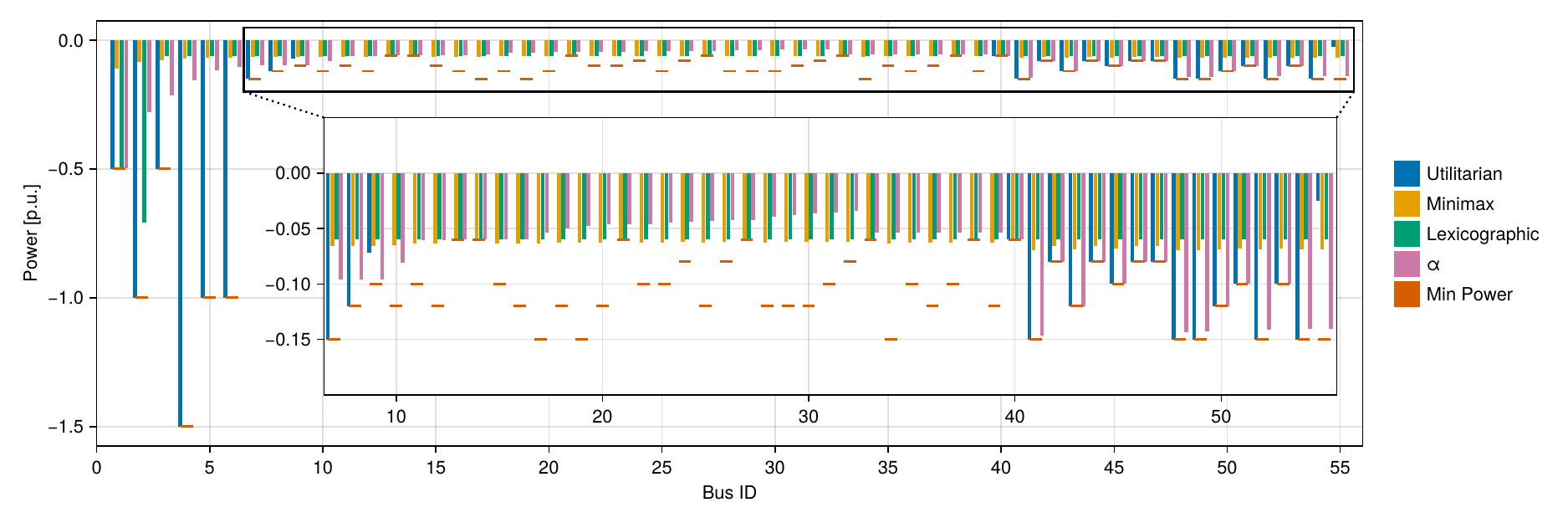}
    } \\
    \vspace{-0.5em}
    
    \subfloat[Maximum and minimum operating envelope for distributed energy resources at each bus.\label{fig:mixed_oe_max_min_barplot}]{
        \includegraphics[width=0.95\linewidth]{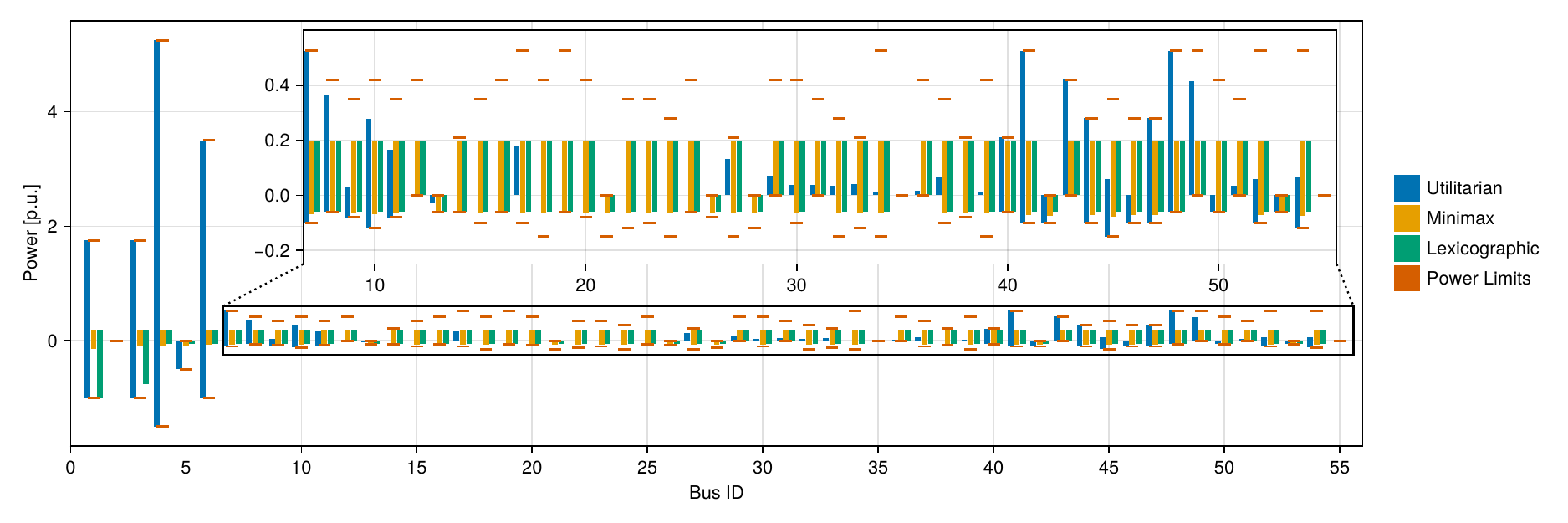}
    }

    \caption{Operating envelope ranges across three scenarios, (a) shows distributed generation at each bus, (b) shows flexible load at each bus, and (c) shows a mixture of some distributed generation and flexible load across most buses in the network.} 
    \label{fig:oe_barplots}
\end{figure*}

\noindent \textbf{Distributed Generation Scenario:}
We first consider the operating envelopes associated with the distributed generation scenario. 
Fig.~\ref{fig:pv_oe_max_barplot} shows the four operating envelope methods applied to the distributed generation scenario. Fig.~\ref{fig:tcl_oe_min_barplot} and Fig.~\ref{fig:mixed_oe_max_min_barplot} show the flexible load and mixed scenarios which will be discussed in later sections. The total size of the DG operating envelopes is in Table~\ref{tab:pv_oe}.  The minimum operating envelope for the DG scenario is 0 for all buses in each problem, as the DG can be curtailed to produce 0 power.

The maximum power of DG resources at each bus is shown with an orange dash.  In blue is the Utilitarian operating envelope. This envelope maximizes the injection potential of resources that are close to the substation, and most other resources are allocated an envelope of size $0$.  Some resources that are electrically close to the substation (40-43 and 47-49, seen in Fig. \ref{fig:pv_scenario}) are allocated an operating envelope. The total capacity is 22.43 p.u.   

\begin{table}[t]
    \centering
    \caption{Total DG Operating Envelope Capacity [p.u.]}
    \vspace{-0.5em}
    \label{tab:pv_oe}
    \begin{tabular}{c|c|c|c|c}
    \toprule
         DER Capacity & Utilitarian & Minimax & Lexicographic & $\boldsymbol{\alpha}=1$ \\
         \midrule
         37.66 & 22.43 & 11.82 & 12.56 & 15.68 \\
        \bottomrule
    \end{tabular}
    \vspace{-1.5em}
\end{table}

In yellow is the Minimax solution, which allocates the same operating envelope to each bus.  The fairness requirement restricts the total amount of operating capacity to 11.82 p.u., just over half of the Utilitarian capacity.  In green is the Lexicographic solution, which is able to capture slightly more capacity than the Minimax solution by increasing the envelope at bus 1.  Finally, the $\alpha$-fairness solution in pink has operating envelopes almost as large as the Minimax fair solutions at most buses, but increases the total amount of capacity from buses electrically close to the substation for a total capacity of 15.68 p.u.. 

\vspace{.2cm}
\noindent \textbf{Flexible Load Scenario:}
Similar results are seen when flexible load is added to the network. Again, the six buses closest to the substation have much larger DERs to model a group of commercial buildings.  The total operating envelope capacity is shown in Table \ref{tab:tcl_oe}.  The pattern of envelope allocation across the network, shown in Fig.~\ref{fig:tcl_oe_min_barplot},  is very similar to that of the DG scenario.
\begin{table}[t]
    \centering
    \caption{Total FL Operating Envelope Capacity [p.u.]}
    \label{tab:tcl_oe}
    \begin{tabular}{c|c|c|c|c}
    \toprule
         DER Capacity & Utilitarian & Minimax & Lexicographic & $\boldsymbol{\alpha}=1$ \\
         \midrule
         10.76 & 7.54 & 3.60 & 4.39 & 4.93 \\
        \bottomrule
    \end{tabular}
\end{table}
The Utilitarian operating envelope enables consumption up to the full capacity of the buses closest to the substation for a total capacity of 7.54 p.u.  The Minimax envelope solution ensures all buses are allowed the same load  for a total of 3.6 p.u.  The Lexicographic envelope solution is able to give buses 1 and 2 additional capacity for a total of 4.39 p.u.  The $\alpha$-fairness solution gives all buses some capacity, but is able to provide some buses a larger envelope for a total capacity of 4.93 p.u.

Both of these scenarios demonstrate how requiring fairness can significantly restrict the total injection capacity of the DERs in the network.  However, not utilizing any fairness metric can result in most buses not being allocated any capacity in their operating envelope.  Next, we explore how fairness could be applied economically while operating with a Utilitarian operating envelope.

\subsection{Economic Fairness}
\noindent \textbf{Distributed Generation Scenario:}
The allocation of economic benefits between the four operating envelopes and five methods of distributing the economic value of the Utilitarian operating envelope are summarized in Table \ref{tab:pv_distributions}.  The first column indicates which method is used to calculate the operating envelope, and the second column indicates how the economic value is distributed. The labels for Minimax and Lexicographic are abbreviated as MM and LG for the Baseline Fair entries.

\begin{figure*}[t]
    \centering
    
    \subfloat[Distributed Generation Scenario\label{fig:pv_dist_value}]{
        \includegraphics[width=\linewidth]{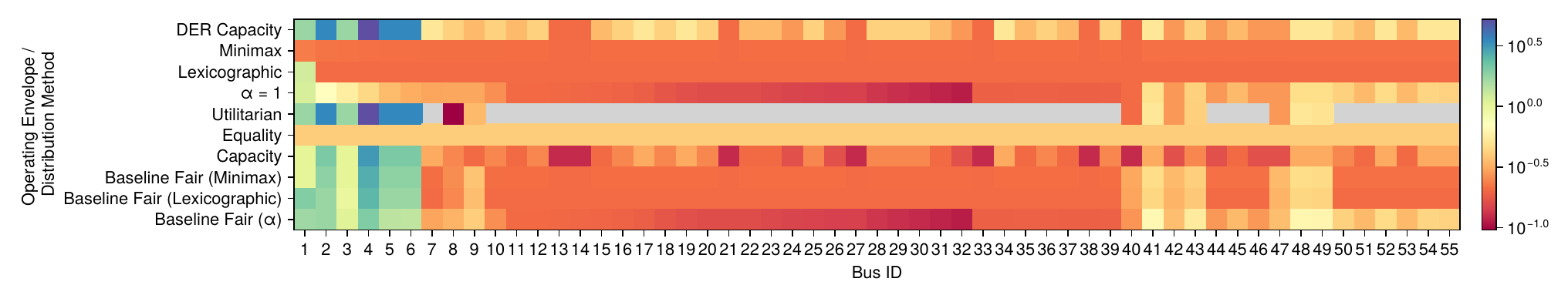}
    } \\
    
    \subfloat[Flexible Load Scenario\label{fig:tcl_dist_value}]{
        \includegraphics[width=\linewidth]{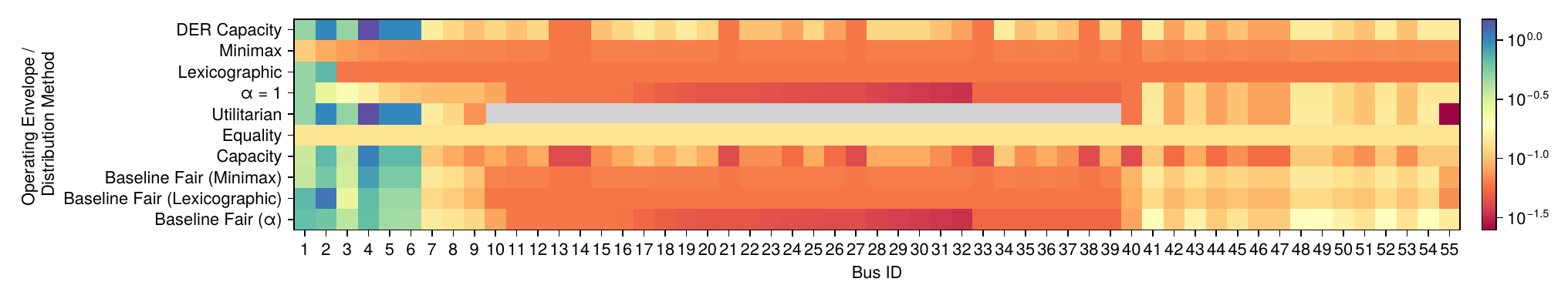}
    }

    \caption{Allocation of value to each bus under the 9 different methods of operating envelope calculation and economic distribution, with the color-scale on a log scale. Grey indicates a value of 0. The first row shows the capacity of the DERs at that bus, while each following row shows the allocated value. The pattern of the results are similar for both the DG scenario in (a) and the FL scenario in (b). The first 6 buses have the largest capacity, and most methods allocate them more value. However, with increasing fairness in operating envelopes the allocation to buses 1-6 decreases. Utilizing economic fairness, with a Utilitarian operating envelope, results in DERs at most of the buses (i.e., 7-55) receiving some compensation even though they are not allocated an operating envelope.}
    \label{fig:dist_value}
\end{figure*}

The first four rows show each of the four operating envelopes with the value allocated according to a Utilitarian distribution.  Among these methods, the total size of the operating envelopes increases as the operations are less fair,  
and using a Utilitarian operating envelope results in many customers being allocated an operating envelope of 0 p.u..  

The last five rows show the effect of different economic distributions while using the Utilitarian operating envelope to physically constrain the DERs.  The total value for each of these distributions is the same because they use the same physical operating envelope, but the distribution of the value varies.

First, the Equality distribution method shares the value equally and this method is able to double the value to each customer compared to utilizing the Minimax operating envelope.  It enables utilization of the full DG capacity of the large facilities near the substation and distributes that value among all members, even when they are allocated an operating envelope that disallows injecting power.

\begin{table}[t]
    \centering
    \caption{DG Allocation of Operating Envelope Value [p.u.]}
    \label{tab:pv_distributions}
    \begin{tabular}{@{}C{5em}C{8.3em}lllll@{}}
        \toprule
Operating Envelope & Distribution Method & Total & Avg & Max & Min & Std \\ \midrule
Minimax & Utilitarian & 11.82& 0.21& 0.23& 0.21& 0.00\\
Lexico & Utilitarian & 12.56& 0.23& 1.22& 0.21& 0.14 \\
$\boldsymbol{\alpha}=1$ & Utilitarian & 15.68 & 0.29 & 1.17 & 0.12 & 0.18 \\
Utilitarian & Utilitarian & 22.43 & 0.41 & 5.25 & 0.00 & 1.08 \\
Utilitarian & Equality & 22.43 & 0.41 & 0.41 & 0.41 & 0.00 \\
Utilitarian & Capacity & 22.43 & 0.41 & 3.13 & 0.13 & 0.58 \\
Utilitarian & Baseline Fair (MM) & 22.43 & 0.41 & 2.70 & 0.21 & 0.51 \\
Utilitarian & Baseline Fair (LG) & 22.43 & 0.41 & 2.52 & 0.21 & 0.51 \\
Utilitarian & Baseline Fair ($\boldsymbol{\alpha}$=1) & 22.43 & 0.41 & 2.03 & 0.12 & 0.44 \\ \bottomrule
    \end{tabular}
\end{table}

The Capacity distribution method distributes the value according to the installed capacity of each DG system.  This means that large systems get paid more, and the smaller systems less, but this also corresponds to their potential contributions to the aggregation's total operating envelope.  

The Baseline Fairness distribution method used in the final three rows ensures that each DER is compensated at minimum the value determined from a reference fair operating envelope problem, and the fairness surplus is allocated proportionally to the physical operating envelopes.  This ensures a baseline level of fairness while compensating the resources according their injection potential.  It is interesting to note here that the Baseline Fair ($\boldsymbol{\alpha}=1$) distribution method is more equal (lower standard deviation) than the Baseline (Minimax) or Baseline (Lexicographic) distribution method because there is a smaller surplus to distribute, despite the fact that the baseline operating envelope values are less equal.

The individual DER economic allocations are shown in Fig. \ref{fig:pv_dist_value}. The first row shows the maximum power capacity of each DG, and each following row corresponds to the row in Table~\ref{tab:pv_distributions}.  This figure clearly shows how the DERs at buses 10-39 receive very small allocations in almost all scenarios, and receive no operating envelope from the Utilitarian method. The Capacity distribution method allocates more value to these DERs by reducing the allocation to DERs at buses 40-55 (which are electrically closer to the substation), but it increases the overall variance among these buses.

\begin{table}[t]
    \centering
    \caption{FL Allocation of Operating Envelope Value [p.u.]}
    \label{tab:tcl_distributions}
    \begin{tabular}{@{}C{5em}C{8.3em}lllll@{}}
        \toprule
Operating Envelope & Distribution Method & Total & Avg & Max & Min & Std \\ \midrule
Minimax & Utilitarian & 3.60& 0.07& 0.10& 0.06& 0.01\\
Lexico & Utilitarian & 4.39& 0.08& 0.71& 0.06& 0.10 \\
$\boldsymbol{\alpha}=1$ & Utilitarian & 4.93 & 0.09 & 0.50 & 0.03 & 0.07 \\
Utilitarian & Utilitarian & 7.54 & 0.14 & 1.50 & 0.00 & 0.30 \\
Utilitarian & Equality & 7.54 & 0.14 & 0.14 & 0.14 & 0.00 \\
Utilitarian & Capacity & 7.54 & 0.14 & 1.05 & 0.04 & 0.20 \\
Utilitarian & Baseline Fair (MM) & 7.54 & 0.14 & 0.85 & 0.06 & 0.16 \\
Utilitarian & Baseline Fair (LG) & 7.54 & 0.14 & 1.13 & 0.06 & 0.20 \\
Utilitarian & Baseline Fair ($\boldsymbol{\alpha}$=1) & 7.54 & 0.14 & 0.67 & 0.03 & 0.16 \\ \bottomrule
    \end{tabular}
\end{table}

\vspace{.2cm}
\noindent \textbf{Flexible Load Scenario:}
The FL scenario results follow the same pattern.  Table \ref{tab:tcl_distributions} summarizes the distribution of values, and Fig. \ref{fig:tcl_dist_value} shows the value allocated to each DER.  
For a flexible load, a limited operating envelope means the flexible load cannot operate at full power to, e.g., maintain thermal comfort or charge an electric vehicle.  This may not matter for a short period of time, but if a flexible load is given too small an operating envelope to provide its energy services (e.g., as a result of operating with a Utilitarian operating envelope), the owner may not care about the economic distributions.

This is evident in Fig. \ref{fig:tcl_dist_value} for the results with Utilitarian operating envelopes but economic fairness.  The Utilitarian operating envelope row shows that DERs connected to buses 10-39 are allocated an operating envelope of size 0, i.e., they cannot operate their flexible load.  Each of the economic fairness methods provide economic value to each of these DERs; however, they are still operated with the Utilitarian operating envelope and will not be able to run.  This can be alleviated by using a more equal operating envelope for operations, like the $\alpha$ fair operating envelope, and still utilizing an economic fairness method to distribution the value, but the issue still stands that some customers might find the value they receive from the aggregator is less than the cost of disruption to their energy services, demonstrating that decoupling physical and economic fairness in the case when the DER is a flexible load may be challenging.

\subsection{Mixed Load and Generation}
The previous sections analyzed two networks that contained only DER load or DER generation.  In this section, we briefly show a network where some buses have generating DERs, some have load DERs, some have both load and generation (and/or storage), and some have neither.   This network uses the same load and generation at each bus as the previous two scenarios, but with some DERs removed.  Buses 2, 5, 13, 21, 26, 28, 35, 42, 53, and 55 do not have generation while buses 2, 12, 29, 31, 25, 36, 43, 49, 51, and 55 do not have flexible load.

Figure \ref{fig:mixed_oe_max_min_barplot} shows the operating envelope limits for a scenario that is the combined flexible load and generation from the previous examples.  The bars in the figure show the total size of the operating envelopes for the Utilitarian envelope, Minimax Fair envelope, and Lexicographic Fair envelope, ranging from net load to net injection.  The $\alpha$ Fair envelope is not shown because of numerical issues when solving for the maximum envelope when a resource has maximum injection of 0 (i.e., no generation), or the minimum envelope when the maximum load consumption is 0 (i.e., no flexible load). This could be addressed with pre-processing to remove these buses from the optimization problem.

The resulting operating envelopes are unsurprisingly similar to the analyzed results above, shown in Figs. \ref{fig:pv_oe_max_barplot} and \ref{fig:tcl_oe_min_barplot}, where buses close to the substation are allocated larger Utilitarian envelopes, and the total size of the envelopes is much larger than the total size of a fair envelopes. This pattern is not as visually recognizable in this network due to the variations in the location of load and generation, but the example demonstrates that this economic fairness approach can be applied to scenarios with heterogeneous resources within the network.

\section{Discussion}
\label{sec:discussion}

There are some very important points to consider when physical fairness is separated from economic fairness.  As seen above, it can allow a significant additional economic value to be gained by the aggregation.  However, there are complications and downsides to consider. We first describe some key considerations for implementing a physically unfair but economically fair operating envelope strategy. We then list some policy recommendations.

\subsection{Key Considerations}

\subsubsection{Resource Types and Services}
The type of DER may impact how much the DER owner cares about physical fairness.  In the case of solar PV, the physical curtailment of a single customer's system may not matter if they are paid fairly by the aggregator.  However, if the DER is an air conditioner, curtailment can have significant impact on the owner.  If the owner is at the end of a feeder and is consistently curtailed so that their neighbors closer to the substation have a increased operating envelope, they likely care very deeply about physical unfairness even if they get a `fair' monetary compensation.  The type of DER and the service it provides to the owner matters, and sometimes physical and economic fairness cannot be easily separated. 

\subsubsection{Incentives to Join or Leave and Aggregation}
DERs may be incentivized or disincentivized to join or leave an aggregation based on the operating envelope approach used, and other factors like the benefit it receives from participating in the aggregation and how often the operating envelopes are active. 
In general, a DER would want to join an aggregation if they would receive sufficient financial incentives from participation, e.g., by providing ancillary services through a virtual power plant. 
Those incentives could change if operating envelopes are often active, reducing the flexibility provided by the aggregation.
Depending upon how the reduction in incentives is distributed across the DERs, an individual DER may continue to benefit, or not, from aggregation, and other DERs may also experience a change in their relative benefits if a DER joins or leaves the aggregation. 

We next demonstrate this through two examples. 
1) Physical and economic unfairness:  a DER joins an aggregation, but is frequently subjected to operating envelope limits.  Therefore, the DER is often not compensated for being a member of the aggregation because it is constrained, despite its intention and ability to participate.   
2) Physical unfairness and economic fairness: a DER joins an aggregation, but is frequently subject to operating envelope limits. Though it is rarely able to contribute to the aggregation, other members are paid less to support the economic fairness of the new member.

\subsubsection{Customer Billing}
Finally, consider the impact of altered DER operations on individual customers' utility bills.  For example, a commercial member may have a demand charge for their largest 15-minute average power consumption each month.  If the aggregator receives a signal to consume more power (stop injecting solar, increase utilization of HVAC) to comply with an ancillary service signal, the commercial customer near a substation may be powering most of the response while other residential members are restricted by network constraints.  All members receive the economic benefits of the aggregator selling flexibility to the ancillary service market, but the commercial member must pay a higher demand charge to the utility.  It is always better for participants in an aggregation to have a contract that jointly manages their electricity/demand payments and incentives to be a member of the aggregation; and the complications of imposing physical and/or economic fairness further emphasize the need for such contracts.

\subsection{Policy Recommendations}

Our formulation and results point to a number of policy recommendations. 

\begin{enumerate}
    \item Decoupling physical and economic fairness shows clear benefits for increasing the amount of rooftop solar that can be injected in a distribution network.  {\em Policies enabling aggregator-coordinated economic fairness could increase the value of rooftop solar and incentivize customers who own solar to the join aggregations.} 
    \item A utility can calculate  Minmax fair operating envelopes across all buses in a distribution feeder and require that DERs follow these limits (e.g., as implemented in Australia). However, an aggregator may prefer alternative allocation of operating envelopes among its members in an unequal way so that it can extract more flexibility from the network.  {\em This would require more coordination and/or data-sharing between utilities and aggregators, which utilities are currently reluctant to do, requiring policy intervention.}
    \item As described in the prior subsection, the DER type and services it provides matter. {\em Policy needs to consider the resource type, rather than treating all DERs as the same, when weighing in on operating envelopes.} Solar and standalone (vs.\ electric vehicle) batteries are more amenable to decoupling of physical and economic fairness, while it makes less sense for flexible loads. One option could be to implement physically unfair distributed generation (injection) operating envelopes, but physically fair flexible load (consumption) operating envelopes. 
    \item Curtailment of DERs (and the need for operating envelopes) typically only occurs infrequently.  When curtailment is not necessary, there is no need to apply fairness either physically or economically to the DER owners/operators.  Any requirements to ensure fairness should only apply to time periods when curtailment in the distribution network is required. \em{Policy is needed to clarify the need and use cases for operating envelopes.}
\end{enumerate}

\section{Conclusion}
\label{sec:conclusion}

This work extends operating envelopes to consider the decoupling of economic and physical fairness.  It describes four different models of computing a physical operating envelope as well as four models of distributing the economic value.  The case study explores the impact of an aggregator using economic fairness and shows that significant value is unlocked by utilizing economic fairness instead of physical fairness.  However, the implementation of this type of approach must consider many non-grid factors, like how a member values their energy services, to correctly incentivize a member to stay in the aggregation.

While this work focuses on a single time period with high levels of congestion, future work can examine the multi-period context and fairness over time.
Additionally, future work will explore the incentives for DER owners to join an aggregation, and the incentive for an aggregator to add members when the aggregator uses economic fairness.

\appendix 
\subsection{AC Optimal Power Flow Equations}
The following equations represent the standard AC power flow equations (see, e.g., \cite{carpentier1962contribution}) used within the operating envelope problems in Section \ref{sec:physical_models}. We present the formulation used in~\cite{coffrin2018powermodels}, which enables computation/optimization of power flow for a generic single-phase equivalent power network. These equations can often be simplified for the radial distribution networks considered in this paper, but we include the full formulation here for completeness.

The substation bus voltage magnitude $V_i$ is held at a fixed value $\hat{\boldsymbol{V}}_i$, and its voltage angle $\theta_i $ is set as the system's reference angle, i.e.,
\begin{subequations}
    \begin{align}
        \theta_i = 0 & \quad \forall i \in \mathcal{B}^\mathrm{ref}, \\
        V_i = \hat{\boldsymbol{V}}_i & \quad \forall i \in \mathcal{B}^\mathrm{ref} .
    \end{align} \label{eq:voltage_reference}
\end{subequations}

Each non-substation bus voltage magnitude $V_i$ is restricted to upper and lower limits, i.e.,
\begin{equation}\boldsymbol{\underline{V}}_i \le V_i \le  \boldsymbol{\overline{V}}_i \quad  \forall i \in \mathcal{B} .\label{eq:ops_voltage_magnitude}
\end{equation}

Thermal apparent power flow limits are enforced on each line, i.e.,
\begin{equation}
    -\boldsymbol{T}_{i j}^2  \le  (P_{i j}^L)^2 + (Q_{i j}^L)^2 \le \boldsymbol{T}_{i j}^2 \quad \forall \, (ij) \in \mathcal{L} . \label{eq:ops_complex_thermal_limit}
\end{equation}

The AC power flow equations are given below.  Parameter $\boldsymbol{g}_{ij}$ is the line conductance, $\boldsymbol{g}_i$ is the shunt conductance, $\boldsymbol{b}_{ij}$ is the line susceptance, and $\boldsymbol{b}_i$ is the shunt susceptance. Parameter $\boldsymbol{t}_{ij}$ is the transformer turns ratio, while $\boldsymbol{t}^R_{ij}$ and $\boldsymbol{t}^I_{ij}$ are the real and imaginary components of the transformer turns ratio.  Then, the active and reactive flow $P^L_{ij}$, $Q^L_{ij}$ are, 
\begin{subequations}
\begin{align}
    \begin{split}
    & P_{ij}^L =   \!\! \Bigg( \!\! \Bigg. 
        \frac{\boldsymbol{g}_{ij} \! + \boldsymbol{g}_{i}}{|\boldsymbol{t}_{ij}|^2} V_i^2 
        + \! \frac{-\boldsymbol{g}_{ij} \boldsymbol{t}^R_{ij} \!+\! \boldsymbol{b}_{ij} \boldsymbol{t}^I_{ij}}{|\boldsymbol{t}_{ij}|^2} V_{i} V_{j} \cos\left(\theta_{i} \!-\! \theta_{j}\right) \\ 
        &  \quad + \! \frac{-\boldsymbol{b}_{ij} \boldsymbol{t}^R_{ij} \!-\! \boldsymbol{g}_{ij} \boldsymbol{t}^I_{ij}}{|\boldsymbol{t}_{ij}|^2} V_{i}  V_{j} \sin\left(\theta_{i} \!-\! \theta_{j} \right) 
    \!\! \Bigg. \Bigg) \: \forall \, (ij) \in \mathcal{L}, \label{eq:ac_p_power_flow_fr}
    \end{split}\\
    \begin{split}
    & P_{ji}^L =  \!\! \Bigg( \!\! \Bigg. 
        \! \left( \boldsymbol{g}_{ij} \!+\! \boldsymbol{g}_{j}\right) \! V_j^2
         \! \! + \! \frac{-\boldsymbol{g}_{ij} \boldsymbol{t}^R_{ij} \!-\! \boldsymbol{b}_{ij} \boldsymbol{t}^I_{ij}}{|\boldsymbol{t}_{ij}|^2} V_{j} V_{i} \cos\left(\theta_{j} \!-\! \theta_{i}\right) \\
        & \quad + \! \frac{-\boldsymbol{b}_{ij} \boldsymbol{t}^R_{ij} \!+\! \boldsymbol{g}_{ij} \boldsymbol{t}^I_{ij}}{|\boldsymbol{t}_{ij}|^2} V_{j}  V_{i} \sin\left(\theta_{j} \!-\! \theta_{i} \right) 
    \!\! \Bigg. \Bigg) \: \forall \, (ij) \in \mathcal{L} , \label{eq:ac_p_power_flow_to}
    \end{split}\\
    \begin{split}
    & Q_{ij}^L = \!\! \Bigg(  \!\! \Bigg.
         \! - \!  \! \frac{\boldsymbol{b}_{ij} \!+\! \boldsymbol{b}_{i}}{|\boldsymbol{t}_{ij}|^2} V_i^2 \!\! 
        - \! \frac{-\boldsymbol{b}_{ij} \boldsymbol{t}^R_{ij} \!-\! \boldsymbol{g}_{ij} \boldsymbol{t}^I_{ij}}{|\boldsymbol{t}_{ij}|^2} V_{i} V_{j} \cos\left(\theta_{i} \!-\! \theta_{j}\right) \\ 
        &  \quad + \! \frac{-\boldsymbol{g}_{ij} \boldsymbol{t}^R_{ij} \!+\! \boldsymbol{b}_{ij} \boldsymbol{t}^I_{ij}}{|\boldsymbol{t}_{ij}|^2} V_{i} V_{j} \sin\left(\theta_{i} \!-\! \theta_{j}\right) 
    \!\! \Bigg. \Bigg) \: \forall \, (ij) \in \mathcal{L} ,\label{eq:ac_q_power_flow_fr} 
    \end{split}\\
    \begin{split}
    & Q_{ji}^L =  \!\! \Bigg( \!\! \Bigg.
         \!- \!\! \left( \boldsymbol{b}_{ij} \!+\! \boldsymbol{b}_{j} \right) \! V_j^2 
         \!\! - \! \frac{-\boldsymbol{b}_{ij} \boldsymbol{t}^R_{ij} \!+\!  \boldsymbol{g}_{ij} \boldsymbol{t}^I_{ij}}{|\boldsymbol{t}_{ij}|^2} V_{j} V_{i} \cos\left(\theta_{j}\!-\!\theta_{i}\right) \!\!\!\! \\ 
        & \quad + \! \frac{-\boldsymbol{g}_{ij} \boldsymbol{t}^R_{ij} \!-\! \boldsymbol{b}_{ij} \boldsymbol{t}^I_{ij}}{|\boldsymbol{t}_{ij}|^2} V_{j} V_{i} \sin\left(\theta_{j}\!-\!\theta_{i}\right) 
    \!\! \Bigg. \Bigg) \: \forall \, (ij) \in \mathcal{L}  \label{eq:ac_q_power_flow_to} .
    \end{split}
\end{align} \label{eq:ac_powerflow}
\end{subequations}

\bibliographystyle{ieeetr}
\bibliography{references}

\end{document}